\documentclass[final,3p,times,authoryear,11pt,review]{elsarticle}

\usepackage{amssymb}
\usepackage{amsmath}
\usepackage{dsfont}

\usepackage{url}
\usepackage[all]{nowidow} 
\usepackage[
    colorlinks= true, 
    linkcolor = blue, 
    urlcolor  = blue, 
    citecolor = blue]{hyperref}
\usepackage{booktabs}

\journal{European Journal of Operational Research}

\DeclareMathOperator*{\argmax}{arg\,max}

\begin{document}

\begin{frontmatter}



\title{Generalizable simulation framework for the request-to-order process in the procurement of onboard vessel requisitions}


\author[inst1,inst2]{Georgios Vassos}

\affiliation[inst1]{organization={Transported by Maersk, A.P. Moller - Maersk},
            addressline={Esplanaden 50}, 
            city={Copenhagen~K},
            postcode={1098}, 
            state={},
            country={Denmark}}

\author[inst2]{Richard Lusby}

\affiliation[inst2]{organization={Department of Technology, Management and Economics, Technical University of Denmark},
            addressline={Akademivej, 358}, 
            city={Kgs.~Lyngby},
            postcode={2800}, 
            state={},
            country={Denmark}}

\author[inst3,inst2]{Pierre Pinson}

\affiliation[inst3]{organization={Dyson School of Design Engineering, Imperial College London},
            addressline={Dyson Building, 1M04A}, 
            city={London},
            postcode={SW7~2AZ}, 
            state={},
            country={United Kingdom}}

\begin{abstract}
    Procurement in maritime logistics faces challenges due to uncertainties in demand and fluctuating market conditions. To address these complexities, we introduce a flexible discrete-event simulation framework that models the request-to-order process. This framework captures critical stages, including the generation of onboard vessel requisitions, requisition handling, and order allocation. Through numerical analysis, we compare two order allocation policies: a naive practice, which relies heavily on contracts, and a dynamic supplier selection approach that explores cost opportunities in the spot market. Our findings reveal trade-offs between cost efficiency and contract compliance, particularly in meeting volume commitments to contracted suppliers. Excessive reliance on spot market opportunities can yield significant savings but at the expense of contract compliance. Additionally, when spot rates are highly sensitive to order quantities, both policies tend to overutilize contracts, highlighting the need for larger volume commitments in such cases. These results offer actionable insights for improving procurement practices, while the framework’s adaptability makes it a powerful decision-support tool across diverse procurement contexts.
\end{abstract}



\begin{keyword}
Simulation \sep  Logistics \sep  OR in maritime industry \sep  Procurement
\MSC 90B06 \sep 90C90 \sep 60G55
\end{keyword}

\end{frontmatter}



\section{Introduction}

Maritime procurement is critical for ensuring the timely and efficient supply of goods necessary for vessel operations. Onboard vessels, crew members regularly submit Purchase Requisitions (PRs) for a wide range of products, including consumables and critical spare parts, to maintain operational readiness during voyages. The Request-To-Order (RTO) process begins with the creation and submission of these requisitions, which are subsequently managed by the Procurement Department (PD). A key step in this process is order allocation, where the items in a PR are assigned to suppliers—either through pre-established contracts or by leveraging spot market opportunities—enabling the issuance of Purchase Orders (POs) to meet supply requirements efficiently and cost-effectively.

In container logistics, where operational continuity is essential, the RTO process supports uninterrupted supply chains and cost-effective procurement, crucial in an industry where inefficiencies can lead to significant operational and financial consequences \citep{Nguyen2023}. For the procurement of onboard vessel requisitions, efficient order allocation is key, involving careful management of lead times and supplier capacities to meet fluctuating demand across multiple ports \citep{Zhou2021}. Given the limited storage onboard, precise inventory control is essential to avoid shortages or excess stock, which could hinder operations. Additionally, the risk of supply disruptions highlights the need for a resilient procurement strategy that ensures continuity while optimizing costs across global shipping routes and variable market conditions \citep{Katsaliaki2022, Nguyen2023}.

Research on the procurement process tends to focus on solving specific problems, such as optimizing costs or improving supplier performance, but often does so without laying the groundwork for a broader, cohesive framework \citep{Chilmon2020}. Though valuable, these studies typically fail to integrate the complex, interdependent factors driving procurement decisions, leading to solutions that may lack generalizability across different contexts and leave broader systemic issues underexplored. A more comprehensive approach is needed to unify these factors and offer strategies better suited to managing procurement complexity.

To address this gap, we introduce a flexible framework that facilitates model customization, expanding its applicability to diverse procurement environments. We develop a conceptual model of the RTO process for onboard vessel requisitions, abstracting its core components into a generalizable ontology. This informs the design of a Discrete-Event Simulation (DES) model with defined events and scheduling logic. Additionally, we formalize the model by parameterizing the RTO process to capture its temporal dynamics, ensuring adaptability across various settings. This approach captures both event timing and data generation, providing a comprehensive solution for modeling the RTO process's evolution.

We conduct a numerical analysis to evaluate the model's internal coherence and its ability to capture key procurement dynamics. Two order allocation policies are assessed: a naive practice, which heavily prioritizes contracts, and a dynamic supplier selection approach, where continuous supplier evaluation informs decision-making. The analysis focuses on two aspects: the timing and nature of vessel requisitions and spot market variations. Assuming all orders are fulfilled either directly from suppliers or via the company’s shipping network—at higher costs due to competition—these assumptions eliminate significant disruption risks, enabling a focused analysis of cost efficiency and contract utilization.

The results indicate that the naive practice overutilizes contracts, thereby missing favorable spot market opportunities, whereas the dynamic policy achieves greater cost savings but risks underutilizing contracts. Although \cite{Dai2020} and \cite{Gur2021} emphasize the trade-offs between contract reliance and spot market flexibility, our work extends this discussion by introducing a quantification of contract compliance, defined as the deviation from initial volume commitments. Additionally, we examine how competition within the spot market influences the effectiveness of both policies.

The remainder of this paper is organized as follows. Section~\ref{sec:lreview} presents a literature review on the significance of supplier selection and order allocation in the procurement process, highlighting the impact of operational constraints, common sources of uncertainty, and key challenges impeding the development of decision-support tools in the maritime logistics industry. In Section~\ref{sec:conceptual}, we develop a conceptual model of the RTO process for procuring onboard vessel requisitions, culminating in a DES model blueprint that illustrates the temporal logic between events. Section~\ref{sec:params} formalizes the time scheduling and event-related data generation components, effectively parameterizing the RTO process. Section~\ref{sec:numerical} demonstrates the practical application of our proposed simulation model through the comparison of two order allocation policies. Finally, in Section~\ref{sec:discussion}, we discuss the importance of our findings, acknowledge the limitations of our study, and propose directions for future research.

\section{Literature review}\label{sec:lreview}

Supplier selection and order allocation are central elements of the procurement process, significantly impacting cost efficiency and overall supply chain performance. Numerous studies highlight the importance of these elements in optimizing costs and balancing multiple objectives like sustainability and operational performance under uncertainty. For instance, \citet{Hamdan2017} and \citet{Wu2023} apply fuzzy decision-making methods---including the Technique for Order Preference by Similarity to Ideal Solution (TOPSIS) and fuzzy numbers---to rank suppliers and allocate orders over multiple periods. These methods address uncertainty by incorporating diverse criteria like sustainability and cost into the decision-making process. In contrast, \citet{Manerba2018} and \citet{Manerba2019} develop stochastic programming models that tackle uncertainties in demand and prices, incorporating factors such as quantity discounts---a reduction in unit cost when purchasing in large volumes---and activation costs, which are the expenses incurred to initiate supplier relationships. Their models provide a quantitative approach to supplier selection and order allocation, emphasizing cost optimization under uncertainty. 

Addressing operational considerations, \citet{Sun2022} propose a model for supplier selection and optimal order-splitting that balances ordering costs, holding costs, and back-order costs. The model accounts for supplier capacity and quality constraints and is validated through discrete-event simulation, demonstrating practical effectiveness. However, as \citet{Saputro2022} distinguish between procurement settings---single vs. multi-sourcing, single vs. multi-item, and single vs. multi-period---they note that most studies focus on decision-making under uncertainty without fully integrating supplier selection with order allocation. This gap suggests a need for models that can jointly address strategic supplier relationships and operational order allocation, as also observed by \citet{Aouadni2019} in their review of recent trends. Our work contributes to this effort by introducing a simulation model of the RTO process that evaluates the performance of order allocation policies and informs strategic decisions by facilitating supplier evaluation based on expected operational performance.

The literature on procurement processes identifies several key sources of uncertainty, which significantly impact supplier selection and order allocation. A major source is demand variability, where fluctuations in market demand influence pricing, order quantities, and inventory management \citep{Dotoli2020, NooriDaryan2019, Park2018}. Supplier performance uncertainty is another critical factor, particularly with regard to quality, reliability, production capacity, and delivery lead times, all of which can deviate from expectations due to unpredictable conditions \citep{Gur2021, NooriDaryan2019, Taherdoost2019}. Price fluctuations in real-time procurement environments further contribute to uncertainty, with spot market prices and logistics costs varying based on supply and demand conditions \citep{Dai2020, Taherdoost2019}. Additionally, in digital procurement platforms, behavioral uncertainty arises from unpredictable user decisions regarding pricing plans and feature adoption \citep{Zutschi2017}. Lastly, broader supply chain risks, such as disruptions, delays, and sustainability challenges, also add complexity, especially in the context of forming long-term partnerships \citep{Wu2023}. These uncertainties collectively underscore the importance of robust, adaptive decision-making frameworks in procurement.

Collectively, these studies recognize supplier selection and order allocation as interconnected yet distinct decision-making problems influenced by various uncertainties. Supplier selection often serves a strategic, long-term purpose---establishing relationships with suppliers that align with the company's goals and can mitigate risks associated with supplier performance and supply chain disruptions. In contrast, order allocation tends to be more operational, focusing on the day-to-day execution of procurement activities and responding to demand variability and price fluctuations. This distinction underscores the need for integrated models that can address both strategic and operational aspects under uncertainty to enhance overall supply chain performance. 

This work proposes a parameterization of the RTO process as the mathematical foundation for developing a DES to evaluate operational order allocation policies. The procurement strategy is embedded within the parameter configuration, encompassing factors, including selected supplier and terms of collaboration (e.g., contract terms or spot market models). This framework facilitates the evaluation of procurement strategies by simulating operational order allocation policies and analyzing various outputs, such as cost and contract compliance, derived from the parameter configuration.

Maritime procurement is inherently complex due to uncertain supplier availability, extended lead times, and fluctuating demand. Unpredictable supplier availability---resulting from varying stock levels and global supply chain disruptions---delays requisition fulfillment \citep{Raza2023}. Lead times are often extended because vessels operate in remote locations, and port operations may be hindered by congestion or infrastructure breakdowns, further complicating procurement timelines \citep{Nguyen2023, Raza2023}. Demand fluctuations add to these challenges, as vessels require a diverse range of products---from consumables to critical spare parts---with inconsistent and unpredictable demand patterns \citep{Raza2023}. These uncertainties, coupled with the global scale of maritime operations, necessitate a robust and adaptive procurement process that efficiently translates requisitions into purchase orders while balancing cost efficiency with operational needs \citep{Raza2023, Zhou2021}.

The RTO process for the procurement of onboard vessel requisitions is characterized by complexities arising primarily from uncertainties in timing, demand, and fluctuating spot market rates. Studies such as those by \cite{Sun2022}, \cite{Wu2023}, and \cite{Zhang2022} underscore the importance of incorporating uncertainty into procurement models through probabilistic approaches. In this context, DES has emerged as a crucial tool for modeling dynamic system behaviors under uncertainty. As noted by \cite{Aouadni2019} and \cite{Chilmon2020}, DES enables the simulation of variable conditions over time, allowing decision-makers to assess and optimize processes more effectively by accounting for the inherent variability in operations. Similarly, \cite{Vieira2020} emphasize DES as a well-established method for capturing the complexity of real-world processes, particularly in supply chains, where intricate interactions and disruptions are prevalent.

The development and adoption of decision-support tools in logistics face several challenges. A key issue is the lack of understanding and expertise in utilizing advanced data analytics, which limits effective implementation across organizations \citep{Nguyen2018}. This is exacerbated by concerns over data security and the fragmented adoption of technologies, where inconsistent integration across different systems prevents seamless data flow \citep{Vieira2020}. The complexity of real-world logistics systems, with their multiple actors and processes, further hinders the application of theoretical models that often fall short in capturing the full scope of operational dynamics \citep{Raza2023}. In the maritime logistics sector, these challenges are intensified by resistance to change, especially in organizations that continue to rely on legacy systems \citep{Raza2023}. Additionally, volatility in the spot market and supply uncertainties make procurement strategies more difficult to optimize, as they must continuously adapt to changing conditions \citep{Zhang2022}. Addressing these limitations requires improved data integration, standardization, and the development of computationally efficient tools capable of real-world validation and adaptation.

In this study, we focus on the procurement of onboard vessel requisitions, also referred to as vessel requisitions or PRs, generated by vessels operated by an intermodal container logistics company. We identify the core steps in the RTO process and detail the activities at each stage. To address gaps in previous research, we propose a holistic DES model that simulates the RTO process, integrating time scheduling, event occurrences, and random variability. Our approach incorporates probabilistic models for timing and event data, parameterizing the RTO process through intensity functions and distributions designed to capture both systemic patterns and external disruptions, such as fluctuating spot market rates and demand uncertainties. This sets the foundation for future research to refine the RTO process, adapt to operational constraints, evolving market conditions, and new data sources, and enhance procurement decision-making.

\section{Problem framework}\label{sec:conceptual}

In this section, we introduce a conceptual model of the RTO process, tailored to the procurement of onboard vessel requisitions. We outline the business context, define our modeling objectives, and propose a foundational blueprint for the DES logic. Our approach follows the guidelines established by \cite{Boyle2022}, ensuring that the model is both contextually grounded and methodologically sound.

\subsection{Request-to-order process}

An intermodal container logistics company operates fleets of vessels on scheduled roundtrips between container ports within its liner shipping network. Each vessel typically has a crew of about twenty members responsible for its operation and maintenance. Consequently, the crew generates recurrent requisitions for products essential to vessel operations, categorized within a multi-layer hierarchy. These requisitions form onboard demand, which is submitted to the PD for allocation to vetted suppliers to ensure reliable fulfillment. 

Fleet-level differentiation in shipping is typically based on vessel size and capacity (e.g., ultra-large vs. feeder vessels), trade routes (e.g., long-haul vs. regional services), vessel type and functionality (e.g., container vessels, feeders, specialized vessels), age and technological generation (modern, energy-efficient vessels vs. older models), sustainability initiatives (such as compliance with emissions standards or use of alteanative fuels), and ownership status (owned vs. chartered vessels). These criteria help optimize fleet management, operational efficiency, and route assignment across global shipping networks \citep{UNCTAD2023}.

\begin{figure}[ht]
    \centering
    \includegraphics[width=\textwidth]{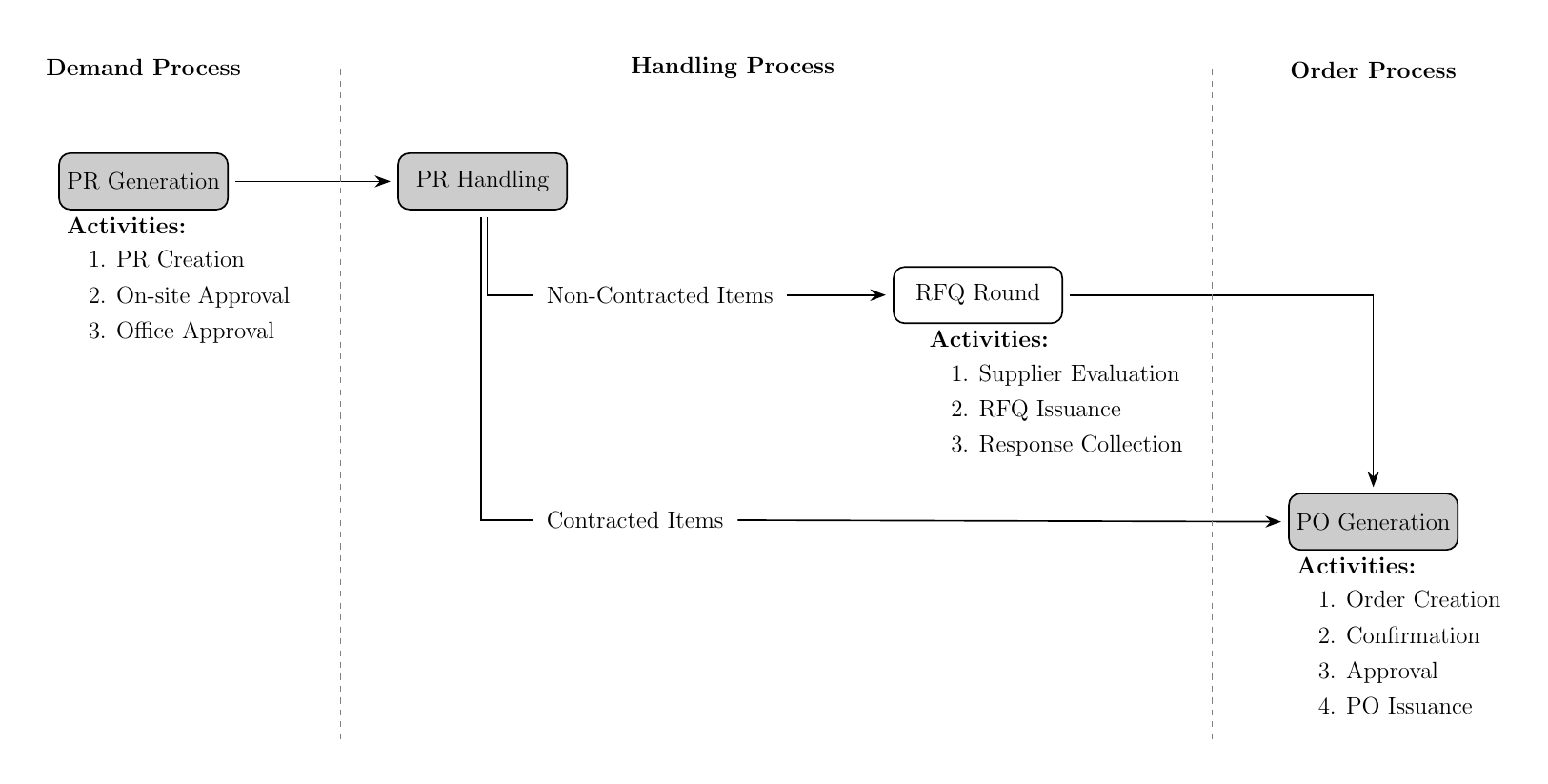}
    \caption{General RTO flowchart for the procurement of onboard vessel requisitions.}
    \label{fig:process_map}
\end{figure}

We decompose the RTO process in the procurement of onboard vessel requisitions into three fundamental subprocesses: the demand process, the handling process, and the order process. As outlined in the previous paragraph, the demand process involves the recurrent generation of category-specific PRs from multiple container vessels. Figure~\ref{fig:process_map} illustrates the general RTO flow for the procurement of PRs.

\subsubsection{Demand process}

In the procurement of onboard vessel requisitions, products are categorized hierarchically to streamline supply management and inventory control \citep{Kendall1986}. Consumables include provisions (e.g., dry, fresh, frozen, beverages), cleaning supplies, personal care items, and shipboard consumables like paints, lubricants, and filters. Spare parts encompass engine and machinery components, electronics, deck and hull equipment, and safety equipment.

The \textit{demand process} is defined as the recurrent generation of PRs from vessels. It is assumed that each PR pertains to a single product category—such as provisions, cleaning supplies, or engine components—potentially to streamline the workflow or simplify coordination. A delay between the creation of a PR and its receipt by the PD may occur, possibly due to administrative or approval procedures.

\subsubsection{Handling process}

The \textit{handling process} of a PR involves identifying all available supply options for each requested item based on the procurement strategy. We define the procurement strategy in terms of two elements: a pre-approved pool of suppliers and a set of contracts. Suppliers are vetted through a strategic selection process, based on a range of quantitative and qualitative criteria such as cost, quality, delivery performance, technological capabilities, service standards, and sustainability criteria \citep{Saputro2022}. Vetted suppliers may be awarded contracts, with transactions conducted either through these contracts or via the spot market. Contracts are valid for a specified period and specify fixed unit prices, supplier commitments on lead times, and company commitments to procure a designated quantity of goods or services during the contract term. The spot market provides a flexible alternative to meet procurement needs when contracted capacities are insufficient.

Contracted items may be immediately forwarded to the order process. For non-contracted items, the PD issues a Request-For-Quotation (RFQ), soliciting bids from multiple suppliers based on required quantities and delivery terms. Through this process, critical information such as pricing and delivery schedules is collected and subsequently passed along to the order process for further execution \citep{Monczka2016, Shokr2017}. In maritime procurement, RFQs are typically sent to pre-approved suppliers evaluated on historical performance, such as reliability and delivery efficiency \citep{vanHoek2022}. In humanitarian logistics, RFQs are often conducted during the announcement phase, inviting bids from suppliers with available inventory and capacity to meet urgent delivery needs \citep{Khoshsirat2021}.

\subsubsection{Order process}

The \textit{order process} depends on information provided by the handling process, specifying whether procurement will follow contract terms or utilize the spot market, and generates POs to be sent to external suppliers accordingly. Each PO is associated with a single supplier, but a single PR may be allocated to multiple suppliers, leading to the creation of multiple POs. Additionally, the generation of POs may incur time delays due to required steps such as confirmation, approval, and issuance before the order is finalized and sent to the supplier.

The core activity of the order process is to determine the allocation of the PR, or parts of it, to suppliers. For contracted items, this allocation must observe the procurement policy, ensuring that the company's volume commitments to each contracted supplier are met by the end of the contract period. In contrast, spot items may be subject to lead time fluctuations due to supplier stockouts, requiring the lead times disclosed in the RFQs to be considered alongside price to ensure both cost efficiency and timely delivery.

\subsubsection{Dynamic supplier selection}

A key issue with the naive practice in the handling and order process is the tendency to excessively utilize the contracts. This could be driven by the misconception that deviating from this approach would breach procurement policy. This often results in contracted suppliers receiving excessive demand volumes, a phenomenon we refer to as \textit{contract overutilization}. Consequently, favorable opportunities in the spot market may be overlooked, which negatively impacts operational efficiency.

To address contract overutilization and enhance operational efficiency, a heuristic method is proposed. This method involves regularly exploring spot market opportunities through RFQs and prioritizing minimum cost over excessive preference of contracts. However, this approach carries the risk of compromising contract compliance, as it may lead to missed volume commitments with long-term partners.

\subsection{Discrete-event simulation model}

We propose a DES model of the RTO process, structured around four events: \emph{PR Generation} (Type 1), which captures the creation and approval of a PR; \emph{PR Handling} (Type 2), where contract data for each item in the PR is retrieved to determine the appropriate procurement path. For contracted items, the process moves directly to \emph{PO Generation} (Type 4), where POs are allocated to suppliers based on a minimum-cost logic and issued following confirmation and approval. For non-contracted items, this triggers an \emph{RFQ Round} (Type 3), during which RFQs are sent to suppliers and responses are collected prompting a PO Generation.

\begin{figure}[ht]
    \centering
    \includegraphics[width = 0.75 \textwidth]{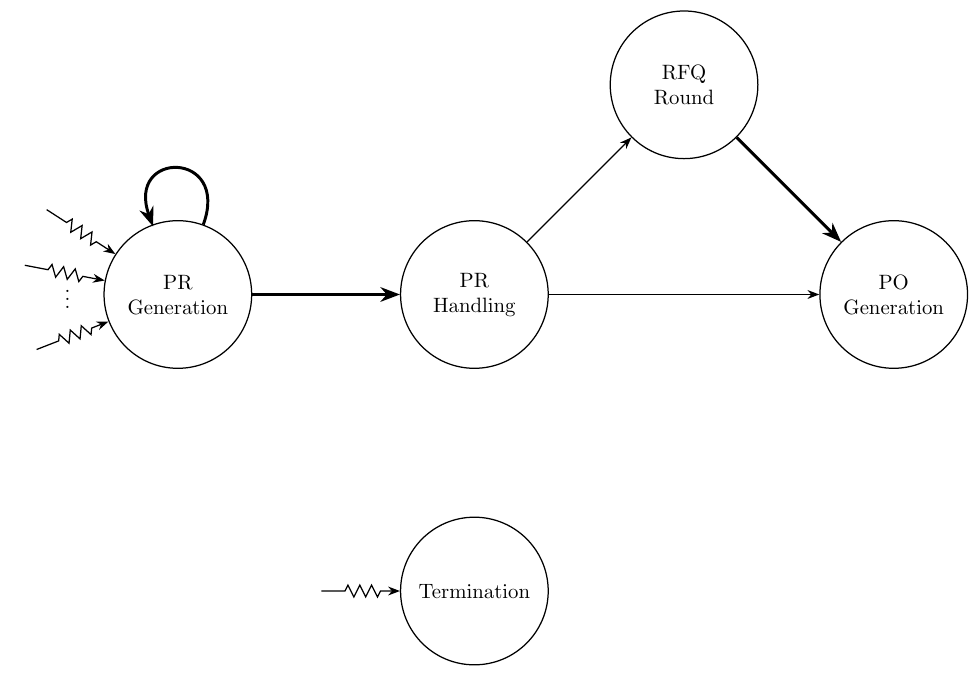}
    \caption{Event graph for the proposed DES model of the RTO process.}
    \label{fig:eg}
\end{figure}

The event-graph method, which represents events as nodes and their scheduling relationships as directed arcs, provides a clear framework for illustrating how events in a system interact and trigger one another \citep{Law2000des}. Figure~\ref{fig:eg} shows our suggested event graph representation of the DES model, illustrating how events are scheduled in the simulation. Jagged arrows indicate event sceduled during initiation, thick arrows represent scheduling with a time delay, and standard arrows denote direct invocations. An auxiliary termination event signifies that the simulation concludes at a predetermined time horizon specified at the start of the simulation.

\subsubsection{Requisition generation}

Vessels generate PRs related to different product categories over a set period. This event simulates the generation of PRs by retrieving past information, determining the likelihood of requesting each product from the relevant category, and generating a request. Future events, such as subsequent requests or the handling of the current PR, are scheduled to keep the process running.

\subsubsection{Requisition handling}

This event retrieves the request details, assesses each supplier’s availability, and verifies whether their contracts are valid for the requested items. Depending on the applied policy, the function can either: (1) forward items from suppliers with valid contracts directly to the order process while invoking an RFQ round for items without contracts, or (2) initiate an RFQ round for all items, regardless of contract status. This approach allows for the implementation of both naive practices and the proposed dynamic supplier selection strategy.

\subsubsection{Request-for-quotation round}

In this event, RFQs are sent to the qualified suppliers identified during strategic supplier selection. Each supplier provides unit prices for the individual items in the request, along with one lead time for fulfilling the entire order. After all bids are received, an order is scheduled, ensuring the procurement progresses based on the submitted quotations.

\subsubsection{Order generation}

In this final event, the requested items are allocated to suppliers according to the procurement policy in effect. Two policies are considered: the naive practice, which prioritizes contracts, and the dynamic supplier selection policy, which considers both contracted and spot market suppliers. The order event assigns items based on the operational policy and issues orders to the suppliers. It also updates procurement costs, accounting for any additional expenses incurred from using multiple suppliers when necessary. Once the allocation is finalized, a PO is generated and stored, concluding the procurement process for that request.

\section{Mathematical model}\label{sec:params}

We consider the RTO process for a fleet of \(n\) vessels, each regularly submitting requisitions for products across \(m\) distinct categories, such as provisions, consumables, or spare parts. Each requisition targets a specific category and includes a list of items and corresponding quantities. These products are organized in a master catalog, \(\mathcal{C}\), which is partitioned into non-overlapping categories \(\mathcal{C}_j\).

The procurement strategy defines a set \( \mathcal{S} \) of vetted suppliers eligible for business transactions. For each category \(j\), a specific subset of suppliers, \(\mathcal{S}_j\), is pre-approved to fulfill requisitions. These suppliers may operate under fixed-term contracts, specifying agreed-upon conditions such as prices, lead times, and volume commitments, or participate in the spot market, where prices and lead times are determined at the time of purchase based on market conditions.

In this model, we aim to formalize the RTO process by capturing the stochastic nature in both timing and data generation along the stages of the process. This framework enables us to simulate the end-to-end flow, assess contract terms and spot market conditions, and analyze order allocation policies in terms of costs and compliance with contractual obligations.

\subsection{Request generation}

Let \( \mathcal{I} \) denote an index set of the vessels. A request or PR is created when a vessel \( i \in \mathcal{I} \) submits a request for products from category \(j\), occurring at a rate governed by an intensity function \(\lambda_{i,j}^{\mathit{PR}}(t)\). The occurrence of PRs over time is modeled by the counting process \(\{N_{i,j}^{\mathit{PR}}(t): t \geq 0\}\), where \(N_{i,j}^{\mathit{PR}}(t) = k\) represents the total number of requests from vessel \(i\) for category \(j\) by time \(t\).

The PR is represented as a vector of quantities and indicators for whether each product is included. Let \( Y(\ell) \in \{0,1\} \) indicate if product \( \ell \) is in the PR, and let \( Q(\ell) \in \mathbb{N} \) denote the quantity requested for \( \ell \). Thus, a PR for products from category $j$ is a vector \( \{(Y(\ell), Q(\ell)) : \ell \in \mathcal{C}_{j}\} \). Each PR is associated with a specific vessel $i$ and category $j$ and is uniquely indexed by the event time \( T_{i,j,k}^{\mathit{PR}} \) at which it was generated.

\subsection{Request handling and RFQ response collection}

Once a PR is generated, its handling is scheduled at a rate specified by the intensity function \(\lambda^{\mathit{HL}}(t)\). The total number of handling events is tracked by the counting process \(\{N^{\mathit{HL}}(t): t \geq 0\}\), where \(N^{\mathit{HL}}(t)\) gives the number of handling events by time \(t\). Depending on the outcome of the handling step, either an RFQ round is initiated or a PO is generated directly. When an RFQ round is invoked for products in category $j$, responses are collected from suppliers \(s \in \mathcal{S}_j\) at a rate described by \(\lambda_s^{\mathit{RFQ}}(t)\). The corresponding counting process \(\{N_s^{\mathit{RFQ}}(t): t \geq 0\}\) tracks the number of RFQ responses collected from supplier \(s\).

To capture the dynamic nature of pricing and lead times in the spot market, distinct from contracted values, we introduce the following notation. Let \( (\ell, s) \) represent product \( \ell \) from supplier \( s \). The contracted price is denoted by \( \bar{r}(\ell, s) \), while the spot market price at time \( t \) is \( R_t(\ell, s) \). Lead times are similarly defined, with \( \bar{w}(s) \) representing the contracted lead time for supplier \( s \) and \( W_t(s) \) representing the spot market lead time at time \( t \). An indicator variable \( C_t(\ell, s) \) specifies whether product \( \ell \) is contracted with supplier \( s \) at time \( t \):
\[
C_t(\ell, s) = \begin{cases} 1 & \text{if product } \ell \text{ is contracted with supplier } s \text{ at time } t, \\ 0 & \text{otherwise}. \end{cases}
\]
The RFQ response, relative to a PR in category $j$, can be formalized as:
\[
\left\{(Y(\ell), Q(\ell), R_{t}(\ell,s), W_{t}(s)) : \ell \in \mathcal{C}_{j}, s \in \mathcal{S}_{j}^{\text{(spot)}}\right\},
\]
where \( \mathcal{S}_{j}^{\text{(spot)}}  \subset \mathcal{S}_{j} \) denotes the set of spot suppliers to whom an RFQ was issued at time $t$ and $\mathcal{S}_{j}$ represents the complete set of suppliers (both spot and contract) capable of providing products in category $j$. To incorporate contract data and the allocation decision \( A(\ell,s) \in \{0, 1\} \) on whether supplier $s$ gets an order that includes item $\ell$ or not, we can extend this representation to:
\begin{equation}\label{eq:mark}
    E(t) = \{(\underbrace{Y(\ell), Q(\ell)}_{\text{PR}}, \underbrace{C_{t}(\ell,s), R_{t}(\ell,s), W_{t}(s)}_{\text{HL, RFQ}}, \underbrace{A(\ell,s)}_{\text{PO}}) : \ell \in \mathcal{C}_{j}, s \in \mathcal{S}_{j}\},
\end{equation}
where, if \( C_{t}(\ell,s) = 1 \), the unit prices and lead time are fixed, with \( R_{t}(\ell,s) = \bar{r}(\ell,s) \) and \( W_{t}(s) = \bar{w}(s) \), corresponding to the contracted values.

We have effectively defined the unit of information, also called the \emph{mark} and denoted by $E(t)$, in Equation~\ref{eq:mark}. Clearly, the components of the mark are progressively collected throughout the RTO steps. The process begins with the generation of a PR, for instance, at time $T_{i,j,k}^{\mathit{PR}}$ for the $k$-th request from vessel $i$ for products in category $j$. Following the PR generation, contract-related data are gathered during the handling of the request, and spot market data are captured during the RFQ round. Finally, the cycle concludes with the PO generation event, where allocation details are finalized.

\subsection{Order Allocation}

Once the RFQ responses are received and evaluated, a PO is generated, occurring at a rate governed by the intensity \(\lambda^{\mathit{PO}}(t)\). The total number of POs issued up to time \(t\) is modeled by the counting process \(\{N^{\mathit{PO}}(t): t \geq 0\}\). Thus, we define the information process \( (\mathcal{F}_{t} : t \ge 0) \) in terms of the \( \sigma \)-field:
\[ 
\mathcal{F}_{t} = \sigma\left\{N^{\mathit{PR}}(u), N^{\mathit{HL}}(u), N^{\mathit{RFQ}}(u), N^{\mathit{PO}}(u), E(u) : 0 \le u < t\right\}, 
\]
where \( N^{\mathit{PR}}(t) = \{ N_{i,j}^{\mathit{PR}}(t) : i=1,\dots,n,\, j=1,\dots,m \} \) for all \( t \ge 0 \).

The PO generation process determines the allocation of requested items to suppliers within the relevant product category. This allocation is guided by a policy $\pi_{t}: \mathcal{E} \times \mathcal{S} \to [0,1]^{|\mathcal{C}_{j}| \times |\mathcal{S}_{j}|}$, where $\mathcal{E}$ represents the set of all possible marks. The policy $\pi_t$ assigns a probability to each supplier for receiving an order for a given combination of products. In some cases, the policy can operate deterministically: if a supplier meets predefined conditions---such as contractual obligations or performance criteria---they are guaranteed to receive the order.

After the allocation is done under the policy, new information is added to indicate whether an order is issued to a supplier $s$ includes item $\ell$ in the $k$-th PR. Let $A_{k}(\ell,s)=\mathds{1}\{Y_{k}(\ell)=1\}\mathds{1}\{s=s_{\ell}^{*}\}$, where $s_{\ell}^{*}=\argmax\{b_{j}(\ell)^{\top}\pi_{t}(s \mid \mathcal{F}_{t}):s\in\mathcal{S}_{j}\}$, where $b_{j}(\ell)$ is a vector of length $|\mathcal{C}_{j}|$ with a 1 at position $\ell$ and zeroes elsewhere.

\subsection{Parameterization of the RTO process}

The formal representation of the RTO process involves two essential aspects, i.e., timing and data generation. The parameters of the timing component are the intensity functions \( (\lambda_{i,j}^{\mathit{PR}}(t):i=1,\dots,n,\,j=1,\dots,m),\ \lambda^{\mathit{HL}}(t),\ (\lambda_{s}^{\mathit{RFQ}}(t):s\in\mathcal{S}),\text{ and }\lambda^{\mathit{PO}}(t) \) that capture the rate at which events occur in our simulation. Let \( \mathcal{G}_{t} \) represent an extended history of the RTO process, encompassing any information beyond that captured by \( \mathcal{F}_{t} \), such that \( \mathcal{F}_{t} \subset \mathcal{G}_{t} \). All intensity processes are \(\mathcal{G}_t\)-predictable, meaning that their evolution depends on the information available up to time \(t\). 

To understand the data generation mechanism, consider the sequence of information associated with the $k$-th request from vessel $i$ for products in category $j$, which unfolds as follows:
\begin{equation*}
\begin{aligned}
    &T_{k}^{\mathit{PR}}, Q_{k}(\ell) : \ell \in \mathcal{C}_{j}\, \wedge \,\{Y_{T_{k}^{\mathit{PR}}}(\ell) = 1\}, \\[3pt]
    &T_{k}^{\mathit{HL}}, (\bar{r}(\ell,s), \bar{w}(s)) : (\ell, s) \in \mathcal{C}_{j} \times \mathcal{S}_{j}\, \wedge \,\{C_{T_{k}^{\mathit{HL}}}(\ell,s) = 1\}, \\[3pt]
    &T_{k,s}^{\mathit{RFQ}} : s \in \mathcal{S}_{j}, \left(R_{T_{k,s}^{\mathit{RFQ}}}(\ell,s), W_{T_{k,s}^{\mathit{RFQ}}}(s)\right) : (\ell, s) \in \mathcal{C}_{j} \times \mathcal{S}_{j}\, \wedge \,\{C_{T_{k}^{\mathit{HL}}}(\ell,s) = 0\}, \\[3pt]
    &T_{k}^{\mathit{PO}}, A_{k}(\ell, s) : (\ell, s) \in \mathcal{C}_{j} \times \mathcal{S}_{j}
\end{aligned}
\end{equation*}
At time \( T_{k}^{\mathit{PR}} \), the request is made, where \( Y_{T_{k}^{\mathit{PR}}}(\ell) = 1 \) indicates whether \( \ell \) is part of the request, and \( Q_{k}(\ell) \) gives the quantity for product \( \ell \). The process then moves to \( T_{k}^{\mathit{HL}} \), where handling takes place. If a contract exists with supplier \( s \) for product \( \ell \) (i.e., \( C_{T_{k}^{\mathit{HL}}}(\ell,s) = 1 \)), the order allocation uses the contract terms \( (\bar{r}(\ell,s), \bar{w}(s)) \), and no RFQ is needed. If no contract is in place (i.e., \( C_{T_{k}^{\mathit{HL}}}(\ell,s) = 0 \)), an RFQ is sent to supplier \( s \) at \( T_{k,s}^{\mathit{RFQ}} \), and their response \( R_{T_{k,s}^{\mathit{RFQ}}}(\ell,s) \) along with supplier information \( W_{T_{k,s}^{\mathit{RFQ}}}(s) \) is collected. Finally, at \( T_{k}^{\mathit{PO}} \), a purchase order is issued, and the allocation \( A_{k}(\ell,s) \) is made, specifying which supplier will fulfill the request for product \( \ell \).

To formalize the data generation mechanism, we distinguish between two key components: the probability density of requests and that governing spot market responses. The generation of a PR involves two sub-decisions: whether an item is included and, if so, at what quantity. We model this using a time-dependent probability density:
\begin{equation}
    p_{t}\{(q(\ell), y(\ell)) : \ell \in \mathcal{C}_{j} \mid \mathcal{G}_{t}\} = \prod_{\ell \in \mathcal{C}_{j}} p_{t}(q(\ell) \mid \mathcal{G}_{t})^{y(\ell)} p_{t}(y(\ell) \mid \mathcal{G}_{t}),
\end{equation}
where \( p_{t}(y(\ell) \mid \mathcal{G}_{t}) \) is the propensity to include product \( \ell \in \mathcal{C}_{j} \) in the request, and \( p_{t}(q(\ell) \mid \mathcal{G}_{t}) \) denotes the probability of requesting \( q(\ell) \) units if \( y(\ell) = 1 \). Both the inclusion decision and the quantity are conditioned on the information set \( \mathcal{G}_{t} \), which evolves over time \( t \).

The second component models the spot market responses in terms of supplier rates and lead times. The probability distribution of these responses is given by:
\begin{equation}
    p_{t}\{(r(\ell,s), w(s)) : (\ell, s) \in \mathcal{C}_{j} \times \mathcal{S}_{j} \mid \mathcal{G}_{t}\} = \prod_{s \in \mathcal{S}_{j}} \prod_{\ell \in \mathcal{C}_{j}} p_{t_{s}}(r(\ell,s), w(s) \mid \mathcal{G}_{t_{s}})^{y(\ell)},
\end{equation}
where \( r(\ell,s) \) represents the rate offered by supplier \( s \) for product \( \ell \), and \( w(s) \) represents the lead time. Both probability densities are conditioned on the supplier-specific information \( \mathcal{G}_{t_{s}} \), which is updated at the time of the RFQ response. The inclusion term \( y(\ell) \) ensures that supplier rates and lead times are only considered for products that were included in the request.

The augmented information \( \mathcal{G}_{t} \) must enable the decoupling of correlations, simplifying the representation into a product form. When the recorded data are insufficient to fully decorrelate item-level or supplier-level instances, latent effects must be considered in addition to capture any unresolved dependencies.

\begin{figure}[ht]
    \centering
    \includegraphics[width=0.95\textwidth]{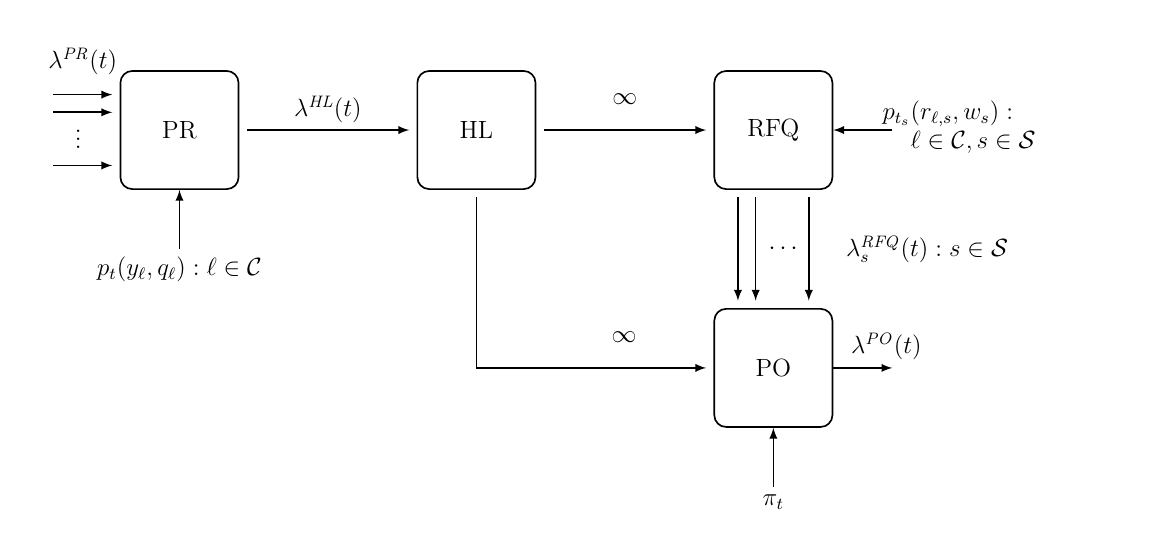}
    \caption{RTO process flowchart showing transitions between stages, with key parameters for time scheduling (intensity functions) and data generation (probability densities) at each step.}
    \label{fig:params}
\end{figure}

Finally, we specify the allocation policy \( \pi_{t} \), which determines how POs are issued based on the cumulative information up to the decision point. Figure~\ref{fig:params} visually aids in understanding the parameterization of the RTO process. Multiple arrows toward the PR represent requests generated by various vessels across product categories, giving rise to events with intensity \( \lambda^{\mathit{PR}}=(\lambda_{i,j}^{\mathit{PR}} : i = 1,\dots, n,\, j = 1, \dots, m) \). Single arrows represent processes such as PR handling or PO issuance, while the RFQ round involves suppliers \( \mathcal{S} \) responding with intensity \( \lambda_{s}^{\mathit{RFQ}} \). This provides a flexible and generalizable parameterization of the RTO process, as depicted in Figure~\ref{fig:process_map}.

Researchers can tailor the proposed framework to their analysis by adjusting the parameters of time scheduling and data generation to suit their specific needs. For those focused on empirical inferences, a rich body of literature offers guidance on fitting and validating flexible models for time scheduling, which is often the most challenging aspect; see, for instance, \citep{Martinussen2006a}. In the following section, we explore how naive order allocation practices perform in terms of cost and contract compliance when compared to the dynamic supplier selection logic under varying spot market conditions.

\section{Numerical experiments}\label{sec:numerical}

We demonstrate the practical application of our DES framework through a hypothetical analysis of cost and compliance outcomes under different order allocation policies. An order allocation policy governs the assignment of items requested in a PR to eligible suppliers. This study compares two specific policies: the naive practice and the proposed dynamic supplier selection approach. We evaluate their performance over a one-year horizon under varying spot market conditions, highlighting the crucial role of these conditions in determining which policy performs better.

Our analysis centers on PR generation, market unit cost conditions, and order allocation policy, employing detailed models for timing and data generation to capture the key dynamics in these processes. In contrast, we simplify the modeling of handling rate, RFQ response rates, and PO generation rate to maintain clarity and avoid unnecessary complexity. This approach allows us to focus on the primary factors that drive cost and compliance outcomes, facilitating a more transparent comparison of the two order allocation policies across different spot market conditions.

\subsection{Demand model}

The generation of onboard vessel requisitions is shaped by the dynamics of internal inventories and the quality levels of products, though the maritime company only observes the timing and content of requisitions, not the inventories themselves. We model the requisition process for a fleet of vessels requesting items within a single product category using a multiplicative intensity function:
\begin{equation}
    \lambda_{i,j}^{\mathit{PR}}(t) = \lambda_{j,0}^{\mathit{PR}}(t - t_{i,j,k}) \exp\{X_{j}(t)^{\top} \beta_{j}\}
\end{equation}
Here, the function \( \lambda_{j,0}^{\mathit{PR}}(t - t_{i,j,k}) \) represents the baseline hazard of a requisition event for products in category \( j \) since the last observed event at \( t_{i,j,k} \), while the term \( \exp\{X_{j}(t)^{\top} \beta_{j}\} \) captures the influence of category-specific time-varying covariates \( X_{j}(t) \). This formulation is flexible, enabling extensions to more comprehensive models that account for demand patterns across heterogeneous fleets and product categories, as in \citep{cook2007a}. For example, a more general specification:
\[
    \lambda_{i,j}^{\mathit{PR}}(t) = \lambda_{i,j,0}^{\mathit{PR}}(t - t_{i,j,k}) \exp\{g_{i,j,t}(X_{i,j}(t))\}
\]

can incorporate vessel- and category-specific baseline hazards, as well as covariate effects \( g_{i,j,t}(X_{i,j}(t)) \). This framework allows for the analysis of variability in requisition generation across different vessels and operational contexts.

To provide a concrete specification, we use a Weibull distribution for the baseline hazard function, which accommodates time-dependent requisition intensity. The Weibull baseline is given by:
\begin{equation}
    \lambda_{j,0}^{\mathit{PR}}(t - t_{i,j,k}) = \frac{\alpha_{1,j}}{\alpha_{2,j}} \left(\frac{t - t_{i,j,k}}{\alpha_{2,j}}\right)^{\alpha_{1,j} - 1},
\end{equation}
where \( \alpha_{1,j} \) and \( \alpha_{2,j} \) represent the shape and scale parameters, respectively. Here, \( \alpha_{1,j} \) governs how the hazard rate evolves with the time since the most recent requisition, and \( \alpha_{2,j} \) sets the relevant time scale. This parameterization provides the flexibility to model increasing or decreasing hazard rates depending on whether \( \alpha_{1,j} \) is greater or less than 1. Alternative baseline hazard specifications are discussed in \citep{cook2007a}, can offer better fits for modeling abrupt changes or exponential trends in hazard rates.

To allow for calendar-driven variations, we can introduce periodic covariates into the multiplicative intensity function. These covariates capture seasonal trends that impact the rate of PR generation. For example, we may include terms like:
\[
\cos\left(\frac{2 \pi t}{\tau}\right) \quad \text{and} \quad \cos\left(\frac{2 \pi t}{\tau} + \frac{\pi}{3}\right),
\]
where \( \tau \) denotes the period of the annual cycle (e.g., 365 days for yearly seasonality). These cosine terms model periodic fluctuations in requisition intensity, potentially reflecting seasonal peaks associated with maintenance schedules or operational cycles.

\begin{figure}
    \centering
    \includegraphics[width = 0.75\textwidth]{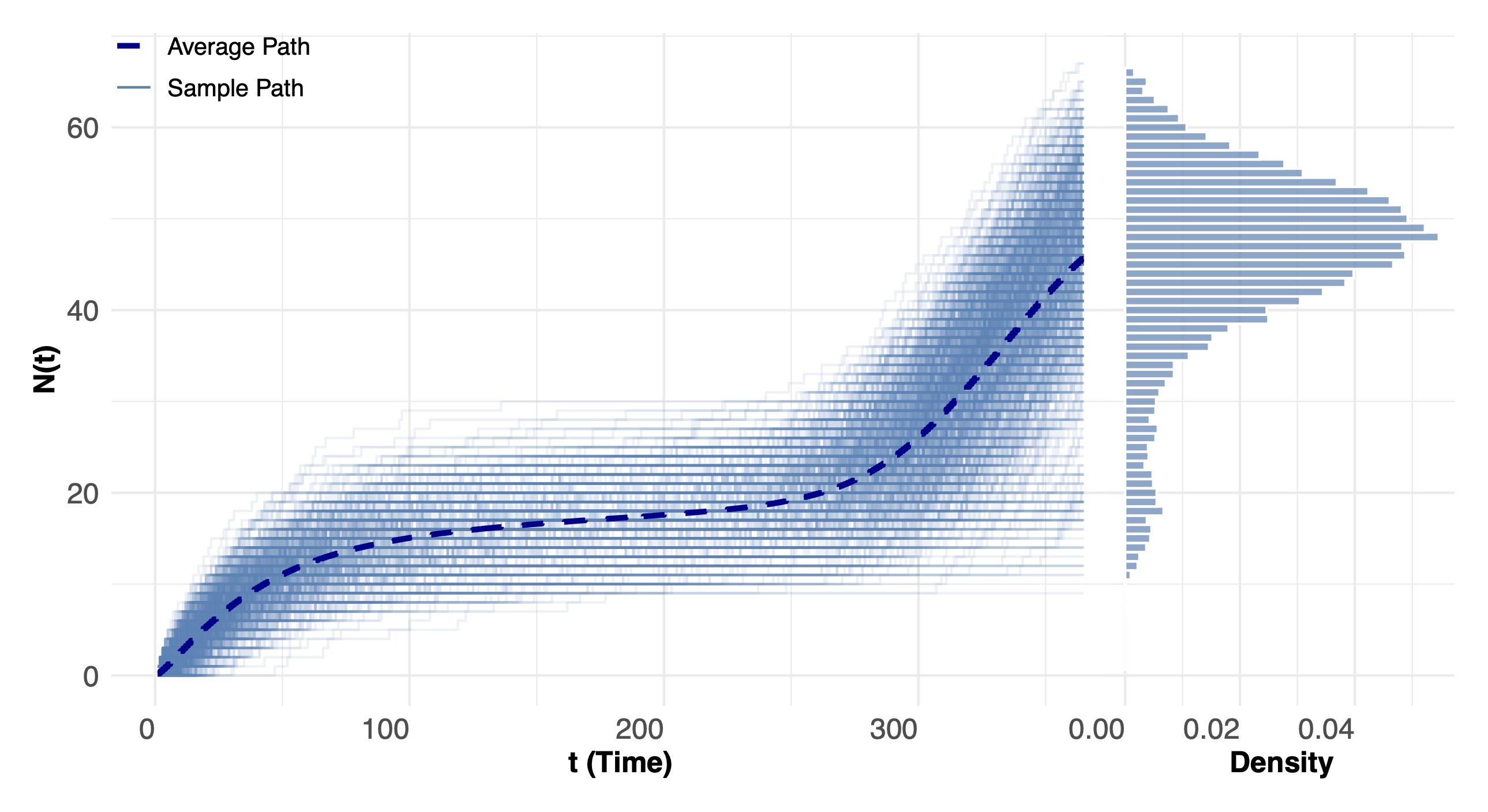}
    \caption{Number of requests accumulated over 365 days from 10,000 simulations (sample paths), with the theoretical cumulative hazard (average path). On the right, a histogram shows the distribution of the total number of requests.}
    \label{fig:demand}
\end{figure}

One challenge in modeling the RTO process for maritime procurement may be the limited visibility into onboard inventory levels and the condition of spare parts. While the crew generates requisitions with complete knowledge of onboard needs, procurement teams can only observe the timing and content of these requests, lacking insights into the underlying dynamics that drive them. To address this limitation, we propose a replenishment model that accounts for unobserved inventory states. Within each category, products are grouped into families, characterized by a baseline stock level and a depletion coefficient, with inventory depleting linearly over time. When an event is triggered by the timing model, the inventory level at that time is computed, and a propensity score is assigned to each product in the category. Items are then sampled for the requisition based on these probabilities, with quantities set to restore inventory to the baseline stock level.

\begin{figure}[ht]
    \centering
    \includegraphics[width = 0.75 \textwidth]{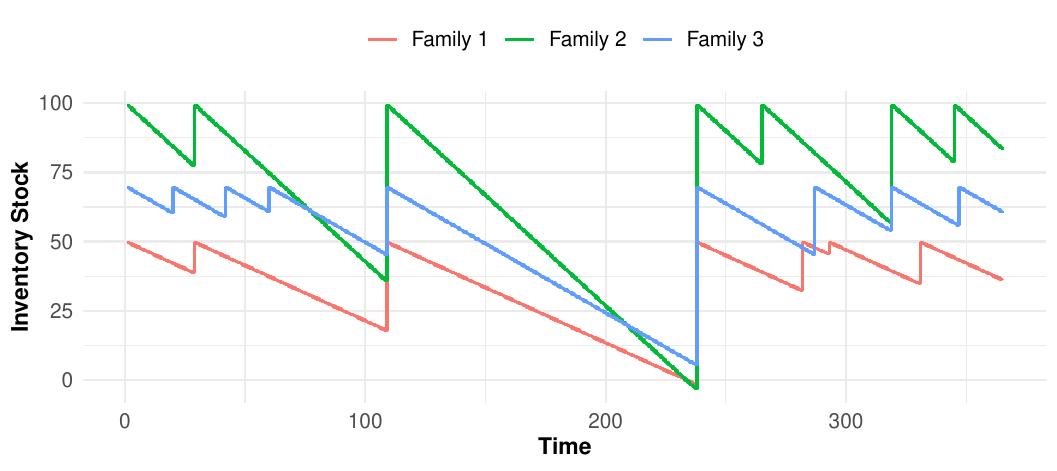}
    \caption{Replenishment cycles for three products from different families within the same category over 365 days, showing depletion and restocking to baseline levels.}
    \label{fig:replenishment}
\end{figure}

Consider products within category \( j \), each associated with a baseline stock level \( q_{0} \) and a depletion rate \( \gamma \in (0,\infty)\). The inventory level for a given product \( \ell \) at time \( t \) is defined as:
\begin{equation}
    q_{\ell}(t) = q_{0} - \gamma (t - t_{\ell}),
\end{equation}
where \( t_{\ell} \) denotes the time of the last replenishment for product \( \ell \). We compute the propensity score \( p_{\ell}(t) \), which reflects the likelihood of including product \( \ell \) in a PR based on the current inventory state:
\begin{equation}
    p_{\ell}(t) = \frac{q_{0} - q_{\ell}(t)}{q_{0}}
\end{equation}
This formulation implies that the propensity score increases as the inventory decreases, indicating a higher likelihood of selecting products with lower current stock levels. The decision to include product \( \ell \) in the requisition at time \( t \) is modeled as a Bernoulli random variable, such that:
\begin{equation}
    Y(\ell) \sim \text{Bernoulli}(p_{\ell}(t))
\end{equation}
If product \( \ell \) is selected for the requisition (\( Y(\ell) = 1 \)), the replenishment quantity \( Q(\ell) \) is determined to restore the inventory to its baseline level:
\begin{equation}
    Q(\ell) = q_{0} - q_{\ell}(t)
\end{equation}
To illustrate this replenishment model, Figure~\ref{fig:replenishment} shows an example of inventory dynamics over 365 days for three products from different families within the same category. Each product's inventory follows a sawtooth pattern, indicating regular depletion at a constant rate and subsequent replenishment to the baseline level. The timing of replenishments varies due to the probabilistic nature of the model, with lower stock levels corresponding to higher propensity scores, thereby increasing the likelihood of replenishment.

Different products have varying consumption rates, leading to distinct inventory management practices, such as frequent restocking for high-turnover items like provisions and less frequent, larger replenishments for durable goods or maintenance supplies. Seasonal demand variations, particularly the need to prepare for winter, drive more aggressive restocking during certain periods \citep{Farhan2018, Katsaliaki2022, Toygar2024}. Additionally, practices like year-end budget utilization and stockpiling before challenging weather or holiday closures influence inventory behaviors. These patterns are also shaped by risk management tailored to the criticality of items, with higher safety stocks maintained for essential supplies \citep{Katsaliaki2022, Toygar2024}.

\subsection{Market model}

The market environment critically influences the performance of procurement policies. We consider contracts with fixed rates, termed strategic rates, for a bundle of products, valid for either six months or one year. These contracts address market uncertainty by securing service at fixed costs, with a commitment to a specified volume of orders from contracted suppliers. The extent to which these agreements are fulfilled during operations defines compliance, reflecting how effectively contracts are utilized.

Ideally, these contracts would cover regular demand, with the spot market serving as a buffer for irregular demand fluctuations. However, initial volume commitments are often rough estimates and may not fully align with actual demand. Additionally, the spot market may offer cost-saving opportunities that were unforeseen at the time of contracting, creating incentives to deviate from commitments. Buyers may also consistently favor contracted suppliers due to their perceived reliability, leading to overutilization of contracts and missed opportunities in the spot market.

We consider a simple spot market environment with three products, each provided by three suppliers. Suppliers A and B have a baseline price of \$10 per unit across all products, while Supplier C has a higher baseline of \$12 per unit. Seasonal trends for each product-supplier pair are modeled as:
\begin{equation}
    \varpi_{\ell,s}\cos\left(\frac{2\pi t}{\tau} + \varphi_{\ell,s}\right), \quad \ell = 1, 2, 3 \text{ and } s = 1, 2, 3
\end{equation}
where \( \varpi_{\ell,s} \) and \( \varphi_{\ell,s} \) are, respectively, the amplitude and initial phase of the seasonal terms, specified in Table~\ref{tab:spotseasonparam}. Random variations are independently drawn from a standard normal distribution for each product-supplier pair and time point: \( \epsilon_{\ell,s,t} \sim \mathcal{N}(0,1) \), for \( \ell = 1, 2, 3 \), \( s = 1, 2, 3 \), and \( t = 1, \dots, \tau \).

\begin{table}[ht]
    \centering
    \small
    \caption{Configuration of parameters $(\varpi_{\ell,s},\varphi_{\ell,s})$ for the seasonal terms.}
    \label{tab:spotseasonparam}
    \begin{tabular}{crrr}
        \toprule
        Product & Supplier A & Supplier B & Supplier C \\
        \midrule
        1 &  \( (2,-\pi/2) \) & \( (3, \pi/2) \) & \( (2,\pi) \) \\
        2 & \( (2,\pi) \) &  \( (3,-\pi/3) \) &  \( (2,\pi/2) \) \\
        3 & \( (2,3\pi/4) \) &  \( (3,\pi/6) \) &  \( (2,2\pi/3) \) \\
        \bottomrule
    \end{tabular}
\end{table}

\begin{figure}[ht]
    \centering
    \includegraphics[width = 0.75 \textwidth]{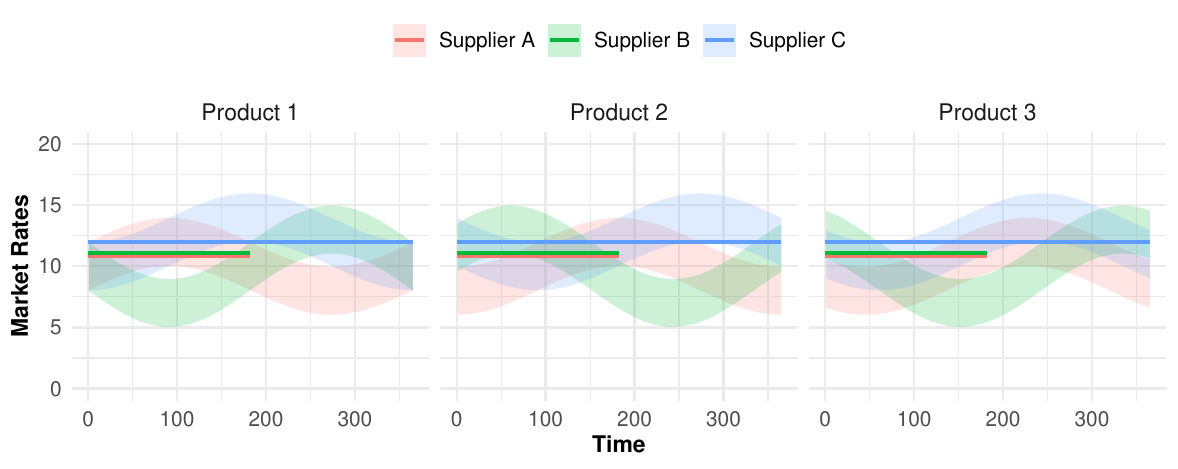}
    \caption{Market conditions for three products from different suppliers. The horizontal lines represent fixed contract rates over their respective time intervals. The harmonic curves show the evolution of spot rates over time for three suppliers. The shaded areas represent the pointwise variability at each time \( t \), assuming a normal distribution of spot rates.}
    \label{fig:MarketConditions}
\end{figure}

Figure~\ref{fig:MarketConditions} illustrates our hypothetical market conditions. The company holds six-month contracts with Suppliers~A and~B, each outlining a fixed unit price of \$11, whereas the one-year contract with Supplier~C specifies a fixed unit price of \$12. Spot market rates for each supplier fluctuate over time, providing cost-saving opportunities. For instance, Supplier A offers more competitive spot rates for Product~1 in the latter half of the year, for Product~2 during the first and last quarters, and for Product~3 in the first quarter and final month.

\subsection{Performance of order allocation policies}

In the previous sections, we introduced the naive order allocation practice and proposed a dynamic supplier selection policy that continuously explores favorable cost opportunities in the spot market. Under naive practice, the allocation of a PR involves listing all contracted suppliers for each item in the requisition and considering the spot market only when no valid contract exists. In such cases, spot market opportunities are identified during the handling process. The minimum-cost allocation is then determined, and orders are placed accordingly. For each extra PO issued, an additional cost of \$10 is incurred due to the administrative overhead associated with processing multiple orders.

In contrast, the proposed dynamic supplier selection policy aims to improve cost efficiency by continuously integrating spot market information. The handling process queries spot suppliers every time a PR is processed, collecting both contract rates and spot market rates for all items in the PR. This ensures that the rates of all qualified suppliers are available before the PR is passed to the order process. During the order process, the minimum-cost allocation is calculated based on the collected rates. As with the naive practice, an additional cost of \$10 is incurred for each extra PO issued.

We compare the performance of the naive practice and the dynamic supplier selection policy across \( 10,000 \) simulation runs, presenting results on both the terminal cost distribution and compliance score. The compliance score is defined as the deviation between the volume allocated to contracts and the initially agreed-upon commitments, effectively capturing contract utilization. In our simulations, the initial agreement specifies 75 units for Suppliers A and B, and 150 units for Supplier C, as the target volumes to be allocated among the suppliers.

In our simulations, the intensities of the handling process \( \lambda^{\mathit{HL}}(t) \), the RFQ response rates of the three suppliers \( (\lambda_{s}^{\mathit{RFQ}}(t) : s = 1, 2, 3) \), and the PO generation process \( \lambda^{\mathit{PO}}(t) \) are all modeled as constants. The total gap time from PR generation to handling consists of two independent exponentially distributed waiting times: one from generation to approval, with a mean of 2 days, and one from approval to handling, with a mean of 5 days. The RFQ response times follow exponential distributions with a mean of 2.5 days for all suppliers. Finally, the waiting time from handling to PO generation is exponentially distributed with a mean of \( 0.1 \) days, for both contracted and RFQ-based orders.

We simulate the RTO model and compare the total cost distributions under two order allocation policies. Our results show that substantial savings can be achieved when the spot market is sufficiently volatile, offering lower-cost alternatives at various times. However, this ideal outcome assumes the spot market can reliably meet all requests without being affected by competition. To assess performance under competitive conditions, we introduce two scenarios where spot rates increase linearly with the quantity requested. In the first, ``mild competition,'' each supplier’s spot rate increases by \$0.01 per unit requested. In the second, ``high competition,'' the spot rate increases by \$0.10 per unit. 

\begin{figure}[ht]
    \centering
    \includegraphics[width = 0.75 \textwidth]{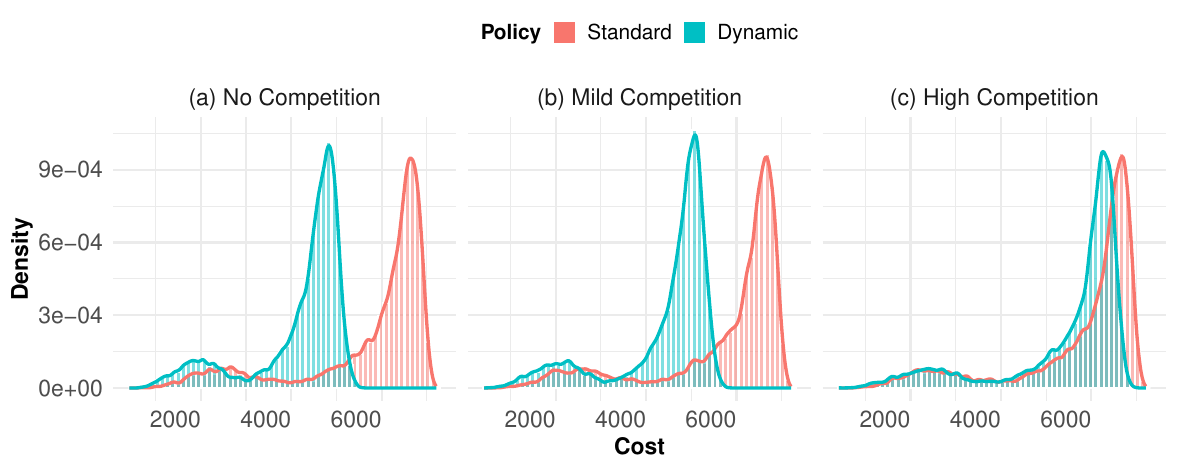}
    \caption{Cost distribution over 10,000 runs under naive and dynamic order allocation policies across three environments: (a) no competition, (b) mild competition (\$0.01 marginal spot rate hike), and (c) high competition (\$0.10 marginal spot rate hike).}
    \label{fig:CostDistr}
\end{figure}

Figure~\ref{fig:CostDistr} presents the cost density under both policies across these scenarios. The cost density exhibits two modes: a lower mode at smaller costs and a higher peak at larger costs, shaped by demand distribution. As competition grows, the cost distribution under the dynamic supplier selection policy progressively aligns with that of the naive practice. This convergence can be attributed to the increasing reliance on fixed contracts as competition in the spot market escalates.

We further analyze contract compliance scores for Suppliers~A, B, and~C across varying competition levels. Under the naive policy, Supplier~C contracts are consistently over-utilized, while Suppliers A and B exhibit variable patterns of underutilization and overutilization depending on demand. In contrast, the dynamic policy displays significant underutilization in the absence of competition, reflecting a preference for spot market opportunities to minimize costs. As competition intensifies, contract utilization adjusts accordingly. Under mild competition, reliance on Supplier~A and~B contracts increases, though the spot market remains the primary source of cost efficiency. However, under high competition, the dynamic policy increasingly leans on fixed contracts as rising spot rates diminish the cost advantage of the spot market.

\begin{figure}[ht]
    \centering
    \includegraphics[width = 0.75\textwidth]{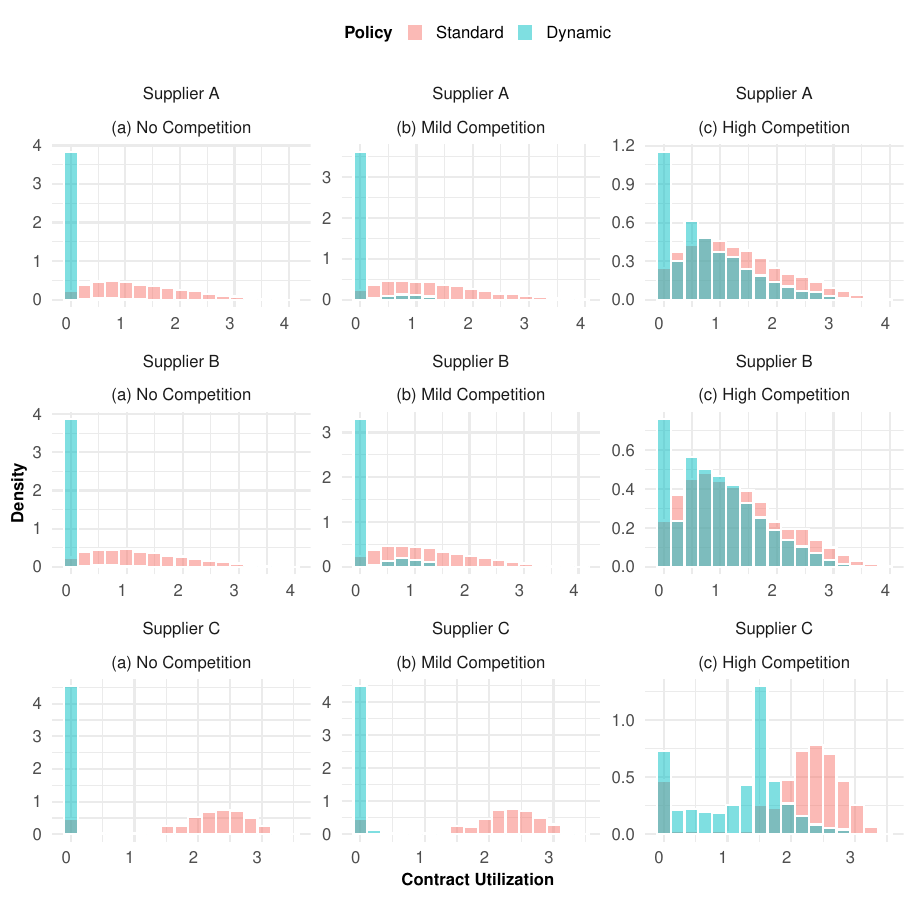}
    \caption{Contract utilization (compliance) distribution across suppliers A, B, and C under naive and dynamic order allocation policies in three environments: (a) no competition, (b) mild competition (\$0.01 marginal cost increase), and (c) high competition (\$0.10 marginal cost increase).}
    \label{fig:ComplDistr}
\end{figure}

Figure~\ref{fig:ComplDistr} illustrates the distribution of contract utilization (compliance) across Suppliers A, B, and C under naive and dynamic order allocation policies in three competitive environments: no competition, mild competition (defined as a \$0.01 marginal cost increase), and high competition (\$0.10 marginal cost increase). The x-axis represents contract utilization, which measures the fraction of a supplier's contract volume commitment that was fulfilled, while the y-axis shows density, indicating the frequency of specific utilization levels.

Under the naive policy, Suppliers~A and~B exhibit right-skewed utilization distributions, with densities peaking at lower values and tapering off at higher values. In contrast, Supplier~C is constantly overutilized across all scenarios. With the dynamic policy, contract utilization varies significantly with the level of competition. In the ``no competition'' and ``mild competition'' scenarios, densities are concentrated near zero for all suppliers, indicating substantial underutilization. In the high competition scenario, utilization shifts toward full or overutilization, with Suppliers~A and~B showing peaks at 0.5 and 0, while Supplier~C has a pronounced peak at 1.5. This pattern reflects the dynamic policy’s response to prohibitive spot market rates, resulting in increased reliance on contracts.

The results demonstrate that high competition creates a strong dependency between spot market rates and order quantities, prompting the dynamic policy to favor contracts, often resulting in overutilization. In contrast, under less competitive conditions, the dynamic policy tends to avoid contracts, leading to reduced compliance. These findings indicate that procurement strategies should prioritize adequate contract volume commitments, especially in highly competitive environments, to achieve a balance between cost efficiency and compliance. This highlights an instance of how the simulation model can be used to evaluate the procurement strategy in terms of the initial volume commitments when agreeing on contract terms with suppliers.

\section{Discussion}\label{sec:discussion}

In academic research on procurement, supplier selection and order allocation emerge as the most frequently addressed challenges. However, there is a growing recognition of the need to approach these two problems in conjunction. We differentiate between these two tasks based on their scope, time horizon, and objectives. Supplier selection is a strategic decision that focuses on identifying suppliers with whom the organization will build long-term relationships, forming the foundation of the procurement process. Beyond supplier selection, the procurement strategy encompasses establishing contracts with fixed rates and terms, engaging the spot market for products not covered by contracts or when contracts expire, and defining the operational order allocation policy that governs how purchase orders are distributed among suppliers.

In contrast, order allocation is an operational process focused on placing orders as demand arises during day-to-day operations. This distinction creates a two-stage framework: the first stage involves formulating the procurement strategy, while the second stage executing it. Performance measures, such as accumulated costs and contract compliance, are used to evaluate the effectiveness of the strategy during its execution. We introduce the concept of contract compliance to measure how well the initial volume commitments, outlined during contract negotiation, are fulfilled. By integrating these stages, strategic decisions can be evaluated and adjusted based on the outcomes observed in operations.

We developed a conceptual framework for the RTO process in the procurement of onboard vessel requisitions and a generalizable DES model by parameterizing the timing and event-related data generation. Given a set of demand and procurement strategy conditions, the simulation resolves demand according to the order allocation policy. At the end of the specified operational horizon, the procurement strategy is evaluated based on metrics such as accumulated costs and contract compliance.

We demonstrated the practical utility of our simulation model through a numerical study, evaluating two operational policies for order allocation. Demand was generated from a small fleet of homogeneous vessels, primarily exhibiting seasonal patterns with a higher intensity of requests during the winter months. The existence of seasonal patters have been reported in the context of maritime logistics by \cite{Toygar2024}, \cite{Katsaliaki2022}, and \cite{Farhan2018}. We further defined the market environment by a set of contracts specifying fixed rates over validity periods, alongside spot rates that exhibit cyclical patterns---such as those arising from periodic supplier stockout---and random volatility. The variability in spot markets has been discussed in \citep{Gur2021, Dai2020}. Our results align with their insights on the cost advantages of regularly exploring the spot market, though excessive exploration may result in underutilization of contracts.

The findings of this study have important implications for both theory and practice. The conceptual map of the RTO process, though focused on onboard vessel requisitions, provides a flexible framework adaptable to other contexts. Translating the RTO into DES logic allows simulating operations to evaluate procurement strategies, assess risks, and measure the impact of disruptions on performance metrics. Future research can leverage this model to explore questions related to optimizing procurement policies, improving resilience, and understanding the interaction between strategic decisions and operational constraints.

This study provides a foundational framework with broad applicability, offering a valuable starting point for addressing diverse research questions in maritime logistics and beyond. While the breadth of the study may not delve deeply into all specific contexts, it allows for flexibility in adapting the framework to other domains, where fewer parameters may simplify implementation compared to the complex maritime setting \citep{Raza2023}. As \citep{Boyle2022} highlighted, focusing on validating the components most relevant to specific research objectives ensures that the framework remains practical and impactful. Although data limitations can influence the fine-tuning of the model with real-world data, they also highlight opportunities for future collaboration with industry stakeholders to enhance accuracy and broaden its applicability.

Future research should focus on advancing the model’s maturity through case studies that extend the complexity of the RTO process to capture intricate demand mechanisms, potentially incorporating heterogeneous product hierarchies, and refining the hazard models with empirical data. Estimating and inferring the model’s parameters to align with real-world complexity remains a critical area of exploration. Another important direction is the development of order allocation policies that minimize costs while achieving optimal compliance. This includes utilizing contracts as closely as possible to initial commitments and efficiently exploring the spot market through predictive models of spot rates and minimizing the number of RFQ interactions required to gather market information.

Additionally, computational scalability requires significant attention, as increasing model complexity—both in terms of parameter count and event routine execution times—may demand extensive computational resources to produce the simulation results needed for analytics. Finally, particular emphasis should be placed on evaluating the impact of disruptions, which are critical in maritime logistics \citep{Nguyen2023} and could provide valuable insights for other industries facing similar challenges.

\section*{Acknowledgements}
This research received funding from the Innovation Fund Denmark. Support from the Danish Maritime Fund has also been applied for / is anticipated / has been granted pending disbursement.

\bibliographystyle{apalike-ejor} 
\bibliography{des-refs}





\end{document}